\documentclass{emulateapj}
\usepackage{amssymb}
\usepackage{graphicx}
\usepackage{amsmath}
\usepackage{natbib}
\usepackage{epsfig}
\usepackage{epstopdf}
\newcommand{\asec} {\mbox{$^{\prime \prime}$} }

\usepackage{mathrsfs}
\usepackage{amsfonts}

\slugcomment{Received to ApJ 2010 July 8; accepted 2010 October 13;
published 2010 October 29} \shorttitle{Hot-Dust-Poor Quasars}
\shortauthors{Hao et al.}

\begin{document}

\title{Hot-Dust-Poor Type 1 Active Galactic Nuclei in the COSMOS Survey}

\author{Heng Hao\altaffilmark{1}, Martin Elvis\altaffilmark{1},
Francesca Civano\altaffilmark{1}, Giorgio
Lanzuisi\altaffilmark{1,2}, Marcella Brusa\altaffilmark{3},
Elisabeta Lusso\altaffilmark{4}, Gianni Zamorani\altaffilmark{4},
Andrea Comastri\altaffilmark{4},  Angela Bongiorno\altaffilmark{3},
Chris D. Impey\altaffilmark{5}, Anton M. Koekemoer\altaffilmark{6},
Emeric Le Floc'h\altaffilmark{7}, Mara Salvato\altaffilmark{8},
David Sanders\altaffilmark{7}, Jonathan R. Trump\altaffilmark{5},
and Cristian Vignali\altaffilmark{9}}

\altaffiltext{1}{Harvard-Smithsonian Center for Astrophysics, 60
Garden Street, Cambridge, MA 02138, USA}

\altaffiltext{2} {INAF-IASF Roma, Via del Fosso del Cavaliere 100,
00133 Roma, Italy}

\altaffiltext{3}{Max Planck Institut f\"ur extraterrestrische Physik
Giessenbachstrasse 1, D--85748 Garching, Germany}

\altaffiltext{4}{INAF-Osservatorio Astronomico di Bologna, via
Ranzani 1, I-40127 Bologna, Italy}

\altaffiltext{5}{Steward Observatory, University of Arizona, 933
North Cherry Avenue, Tucson, AZ 85721, USA}

\altaffiltext{6}{Space Telescope Science Institute, 3700 San Martin
Drive, Baltimore, MD 21218, USA}

\altaffiltext{7}{Institute for Astronomy, University of Hawaii, 2680
Woodlawn Drive, Honolulu, HI 96822, USA}

\altaffiltext{8}{IPP-Max-Planck-Institute for Plasma Physics,
Boltzmannstrasse 2, D-85748, Garching, Germany}

\altaffiltext{9}{Dipartimento di Astronomia, Universit\`a degli
Studi di Bologna, via Ranzani 1, I-40127 Bologna, Italy}

\email{hhao@cfa.harvard.edu, elvis@cfa.harvard.edu}

\begin{abstract}
We report a sizable class of type 1 active galactic nuclei (AGNs)
with unusually weak near-infrared (1--3$\mu$m) emission in the
XMM-COSMOS type 1 AGN sample. The fraction of these
``hot-dust-poor'' AGNs increases with redshift from 6\% at low
redshift ($z<2$) to 20\% at moderate high redshift ($2<z<3.5$).
There is no clear trend of the fraction with other parameters:
bolometric luminosity, Eddington ratio, black hole mass and X-ray
luminosity. The $3\mu$m emission relative to the $1\mu$m emission is
a factor of 2--4 smaller than the typical Elvis et al. AGN spectral
energy distribution (SED), which indicates a `torus' covering factor
of 2\%--29\%, a factor of 3--40 smaller than required by unified
models. The weak hot dust emission seems to expose an extension of
the accretion disk continuum in some of the source SEDs. We estimate
the outer edge of their accretion disks to lie at
(0.3--2.0)$\times10^4$ Schwarzschild radii, $\sim$10--23 times the
gravitational stability radii. Formation scenarios for these sources
are discussed.
\end{abstract}


\section{Introduction}
Characteristically, active galactic nuclei (AGNs) have hotter dust
than starburst galaxies, which has long been employed to select AGNs
in near-infrared (NIR) surveys (e.g., Miley et al. 1985, Lacy et al.
2004, 2007, Stern et al. 2005, Donley et al. 2008). In AGNs, dust
reaches maximum temperature ($\sim$1000--1900 K) at the smallest
equilibrium radius (Barvainis 1987, Suganuma et al. 2006). The hot
dust is assigned to the inner edge of the `torus' by the unified
model of AGNs (Krolik \& Begelman 1988, Antonucci 1993, Urry \&
Padovani 1995). The obscuring torus might be a smooth continuation
of the broad-line region (BLR; Elitzur \& Ho 2009), with the BLR
extending outward to the inner boundary of the dusty torus (Suganuma
et al. 2006).

However, exceptions have been observed and predicted.
Observationally, Jiang et al. (2010) found two $z\sim 6$ quasars,
without any detectable emission from hot dust, suggesting that these
hot-dust-free AGNs are the first generation quasars that do not have
enough time to form a dusty torus. Rowan-Robinson et al. (2009)
found AGNs with torus to bolometric luminosity ratio of only a few
percent.

Emmering et al. (1992) proposed a BLR/torus structure in the
disk--wind scenario. The two structures correspond to different
regions of a clumpy wind coming off the accretion disk rotating
around the black hole (BH). The disk--wind scenario predicts that
the torus disappears at luminosities lower than $\sim 10^{42}$ erg
s$^{-1}$ because the accretion onto the central BH can no longer
sustain the required cloud outflow rate \citep{elitzur06}. In a
sample of nearby AGNs, even the BLR disappears at luminosities lower
than $5\times10^{39}(M/10^7M_{\bigodot})^{2/3} $erg s$^{-1}$
\citep{elitzur09}.

The AGN structure is reflected in the shape of the spectral energy
distribution (SED). The maximum dust temperature leads to a
characteristic drop in the NIR emission at 1$\mu$m (Elvis et al.
1994, E94 hereafter; Glikman et al. 2006). 
AGNs with weak or no `torus' have an SED with weak or no IR bump, or
no 1$\mu$m inflection.

We have found a substantial ($\sim$10\%) population of such
``hot-dust-poor'' (HDP) AGNs by studying the SEDs of 408 XMM-COSMOS
(Hasinger et al. 2007; Cappelluti et al. 2009) X-ray-selected type 1
(FWHM $>$ 2000 km s$^{-1}$; Elvis et al. 2010) AGNs. All these X-ray
sources have secure optical and infrared identifications (Brusa et
al. 2010) and at least one optical spectrum, from either the
Magellan (Trump et al. 2009a), Sloan Digital Sky Survey (SDSS;
Schneider et al. 2007), or Very Large Telescope (VLT; Lilly et al.
2007, 2009) surveys.

\section{Selection of HDP Objects}
The HDP AGNs are selected based on the NIR and optical SED shapes.
We plot NIR versus optical slopes on either side of $1\mu$m (rest
frame, Figure 1; for details see Hao et al. 2010). The 1$\mu$m point
is not chosen arbitrarily. It is where the blackbody emission of the
hot dust at the maximum sublimation temperature (e.g., 1500 K;
Barvainis 1987) normally begins to outshine the emission of the
accretion disk.

The plot is equivalent to a color--color plot, but utilizes more
than four photometric bands. Briefly, we fit power laws ($\nu
F_{\nu}\propto \nu ^{\alpha}$) on either side of the 1$\mu$m
inflection point of the rest-frame SED: $1\mu$m--3000\AA\ to derive
an optical slope ($\alpha_{OPT}$) and $3\mu$m - 1$\mu$m to derive a
NIR slope ($\alpha_{NIR}$). In normal type 1 AGNs, these slopes are
$\alpha_{OPT}\sim1$ and $\alpha_{NIR}\sim-0.7$ (E94; Richards et al.
2006). In order to obtain slopes that reflect the shape of the SED
correctly, we required the number of photometric points of the
linear fit to be larger than 2. Four hundred four out of the 408
X-ray-selected type 1 AGNs\footnote{Three out of the other four
(XID=2119, 5320, 5617) are low-redshift AGNs with only $K$ band and
one IRAC band in the rest-frame NIR SED. The remaining source
(XID=54439) is the highest redshift AGN with only two IRAC bands in
the rest-frame NIR SED.} satisfy these criteria. The errors of the
slopes are the standard errors of the linear fit.

The $\alpha_{OPT}-\alpha_{NIR}$ plot can be conveniently used to
separate the nuclear from the host galaxy emission. Figure 1 shows
the $\alpha_{OPT}$ and $\alpha_{NIR}$ for the E94 radio-quiet AGN
template (red cross) and three galaxy templates (a spiral --- Spi4,
a 5 Gyr old elliptical --- Ell5, and a starburst --- NGC 6090; big
blue triangles) from the ``SWIRE Template Library" (Polletta et al.
2007). The black curves show the slopes of the SED templates
obtained by mixing the AGN and galaxy with different fraction
(0\%--100\%) after normalizing the E94 and galaxy templates at
$1\mu$m. The mixing curves of NGC 6090 and Spi4 define the
boundaries of the possible slopes obtained by mixing the E94 with
all the 16 available galaxy templates in the SWIRE library (small
blue triangles). The magenta line shows the OPT and NIR slopes of
E94 template when reddening ($E(B-V)=0-1$ mag) is
applied\footnote{We applied the IDL dereddening routine
\texttt{``FM\_UNRED.PRO''}, with the SMC extinction curve (Gordon et
al. 2003).}. Reddening primarily affects $\alpha_{OPT}$.

The majority of the XMM-COSMOS AGNs ($\sim$ 90\%) are explained by
combinations of an AGN, a galaxy, and reddening (Hao et al. 2010),
as they lie within the triangular region at the left of the
$\alpha_{OPT}-\alpha_{NIR}$ plot, bounded by the reddening and
mixing curves. However, $\sim$~10\% (41 sources) lie in the upper
right corner of Figure~1, more than $1\sigma$ beyond the
AGN--host--reddening triangle. These sources have an optical slope
consistent with typical AGNs, but a relatively weak IR bump. We name
them ``hot-dust-poor'' (HDP) quasars.

\begin{figure}
\epsfig{file=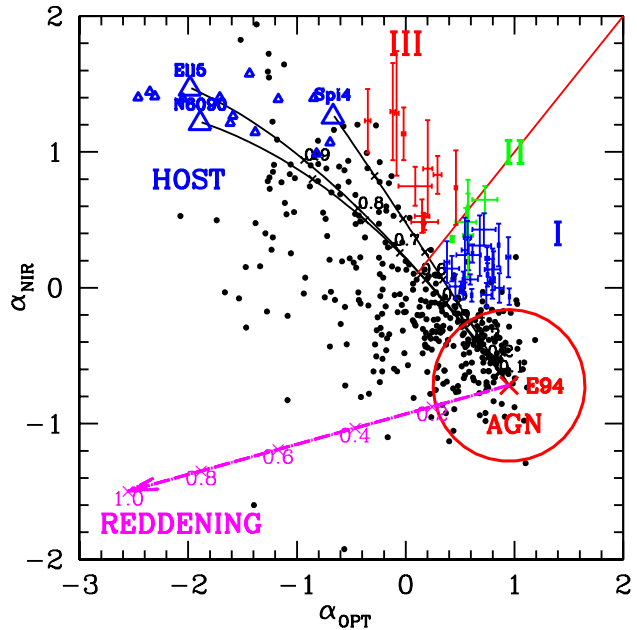, angle=0,width=\linewidth} \caption{Slope--slope
plot of the XMM-COSMOS type 1 AGNs. Red cross and red circle show
the E94 mean SED and the $1\sigma$ dispersion of the E94 sources.
The blue triangles show different SWIRE galaxy templates. The black
lines connecting the E94 and the galaxy templates are mixing curves.
The purple line shows the reddening vector of E94. The straight red
solid line shows the $\alpha_{OPT}=\alpha_{NIR}$ line. Different
colors of the points show different class of the HDP AGNs (I, II,
and III, see the text for details). The black dots show all the
other XMM-COSMOS type 1 AGNs.\label{slpsel}}
\end{figure}

\section {HDP SEDs}
There is a range of SED shapes for the 41 HDP AGNs. We further
divide these AGNs into three classes, according to their relative
positions to the equal slope (red) line in the
$\alpha_{OPT}-\alpha_{NIR}$ plot. Figure 2 shows examples of SEDs
for each class. All wavelengths in this section refer to the rest
frame.

\begin{figure*}
\includegraphics[angle=0,width=0.325\textwidth]{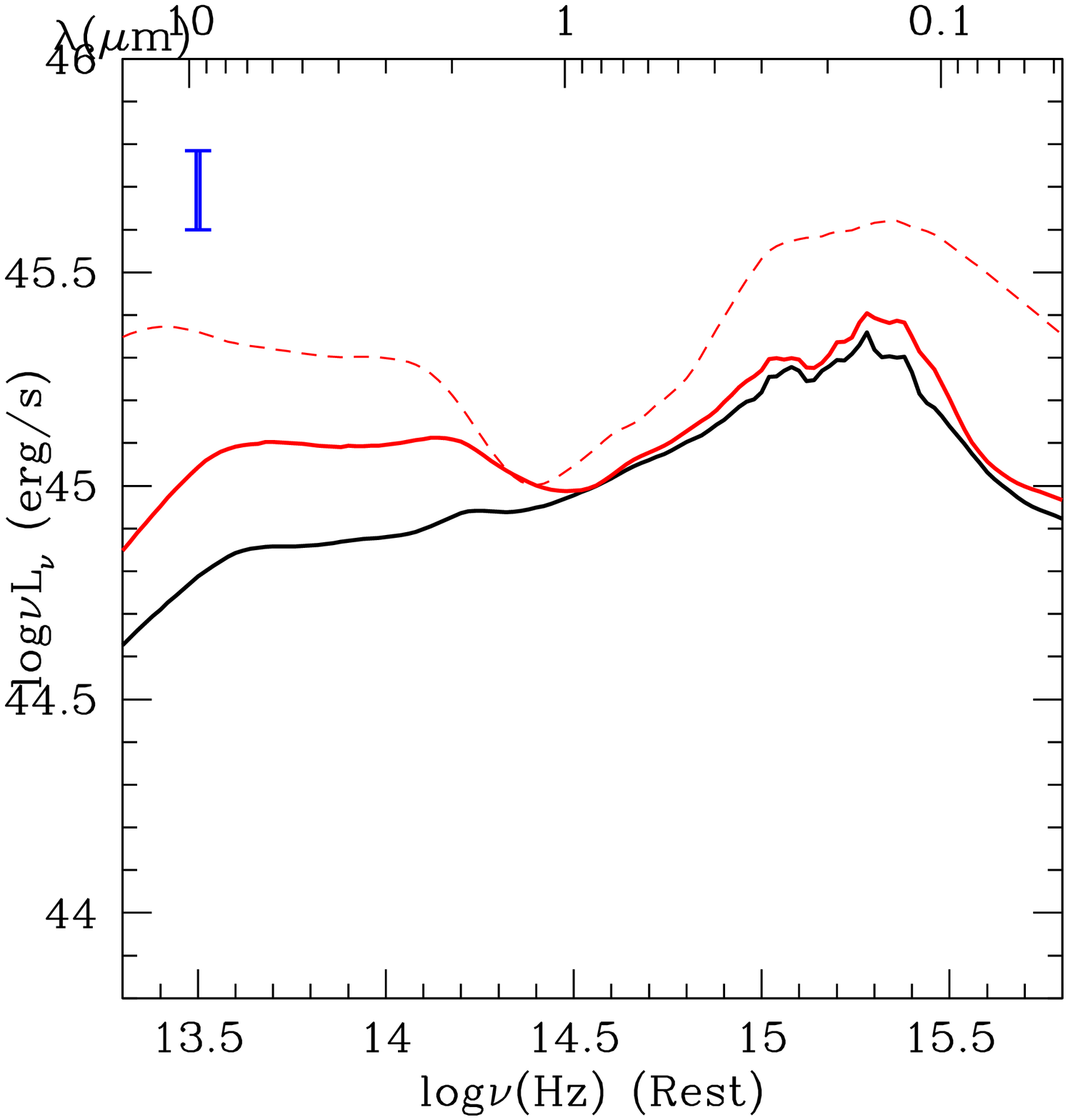}
\includegraphics[angle=0,width=0.325\textwidth]{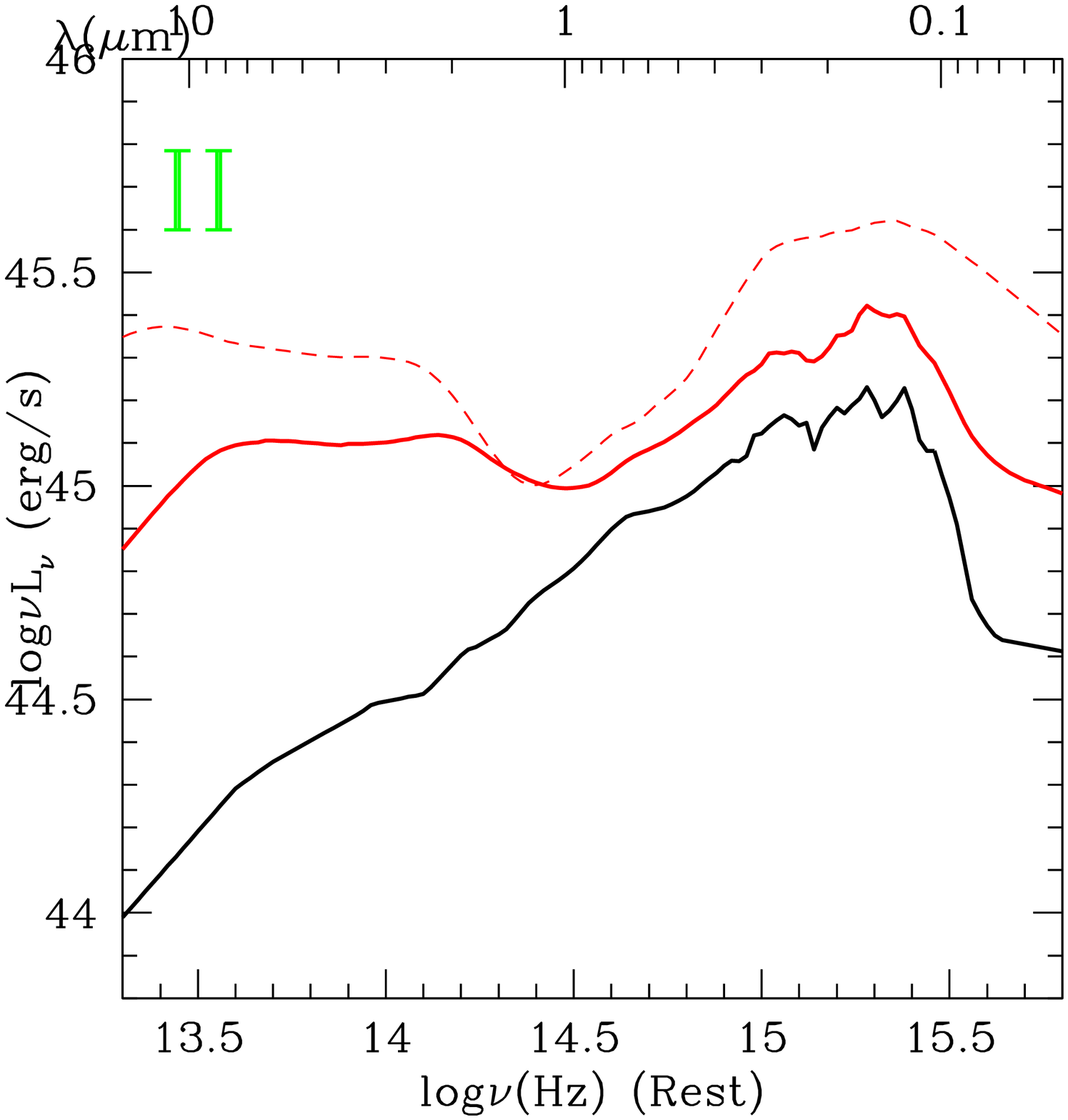}
\includegraphics[angle=0,width=0.325\textwidth]{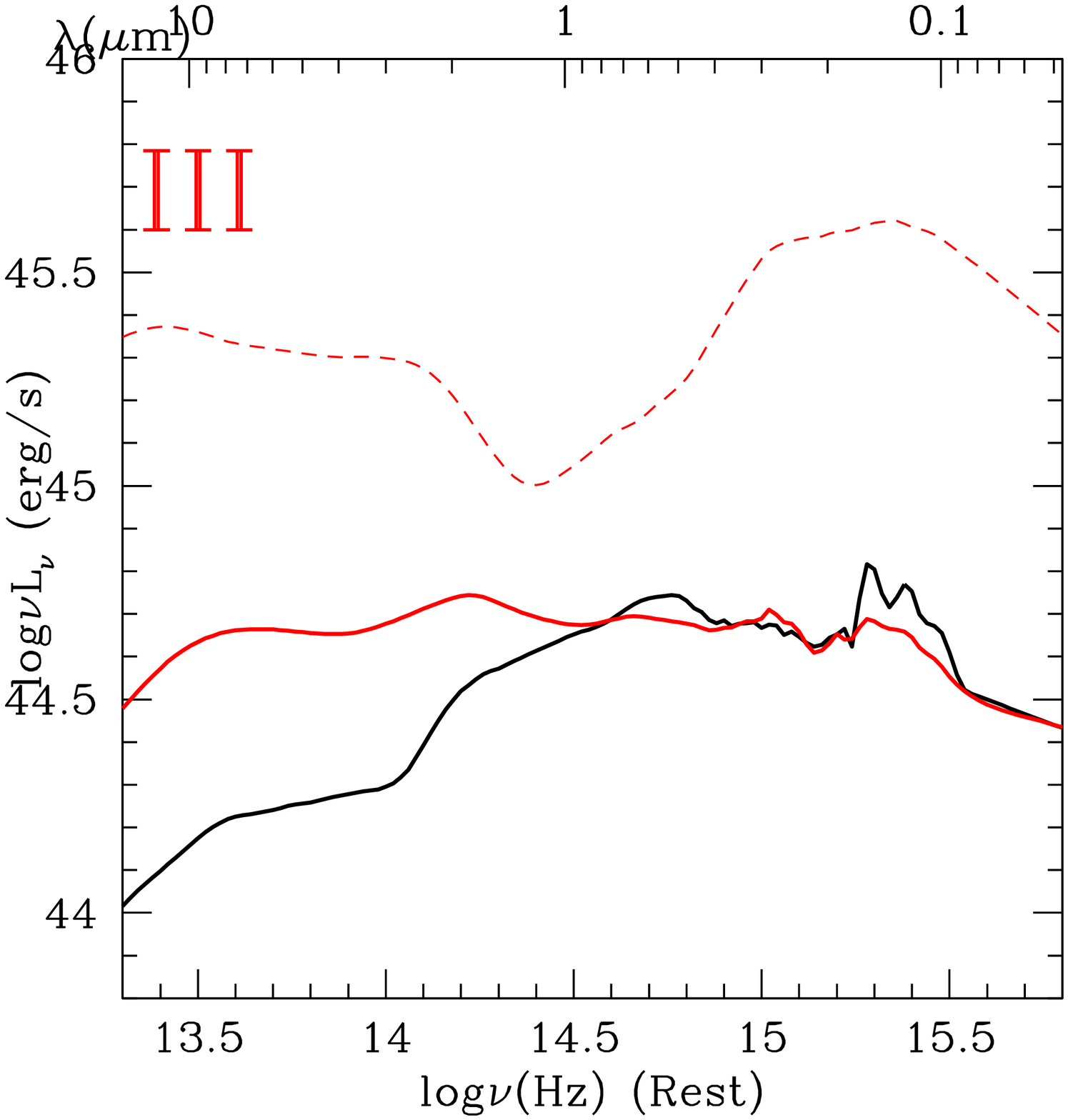}
\includegraphics[angle=0,width=0.325\textwidth]{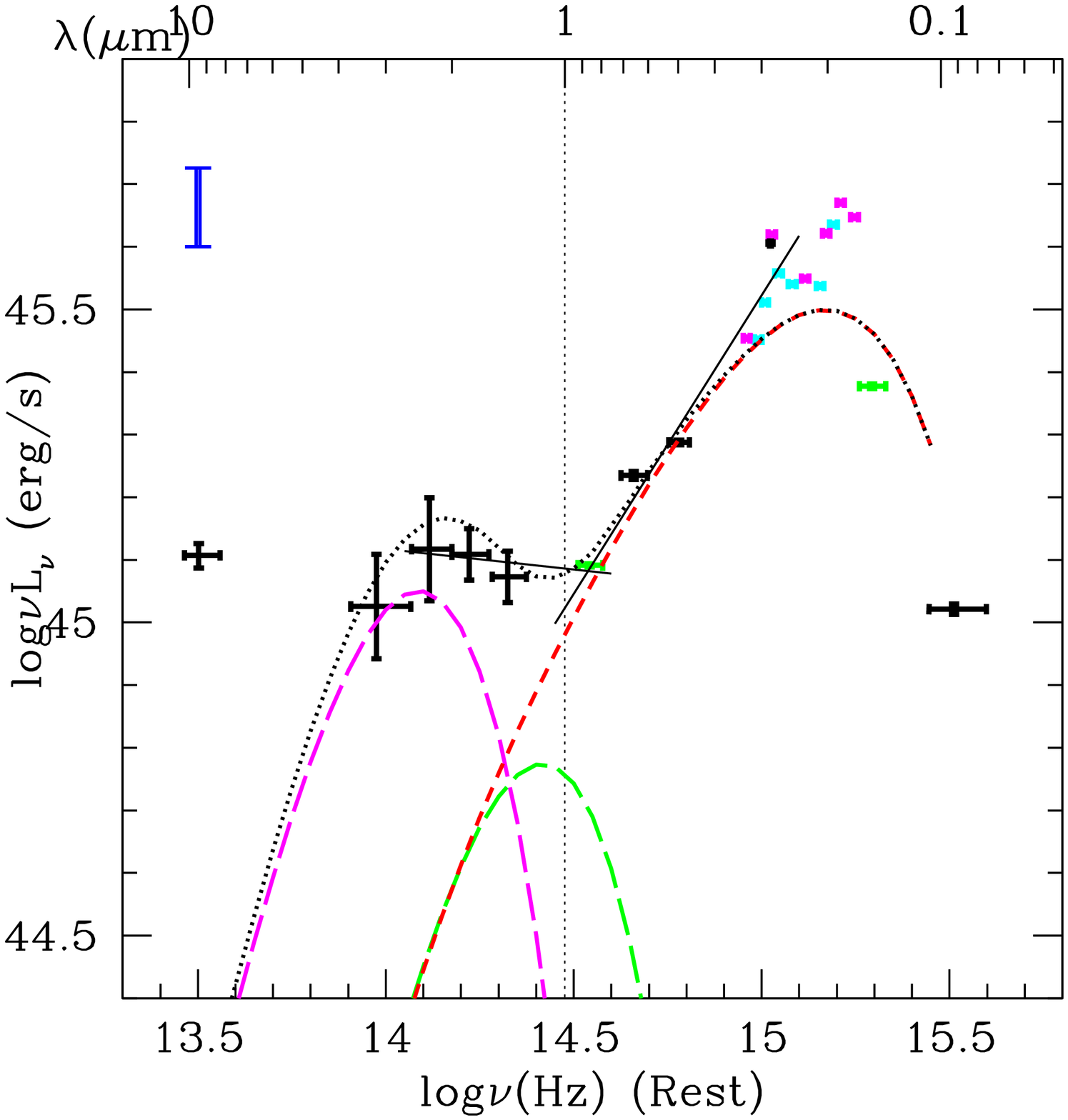}
\includegraphics[angle=0,width=0.325\textwidth]{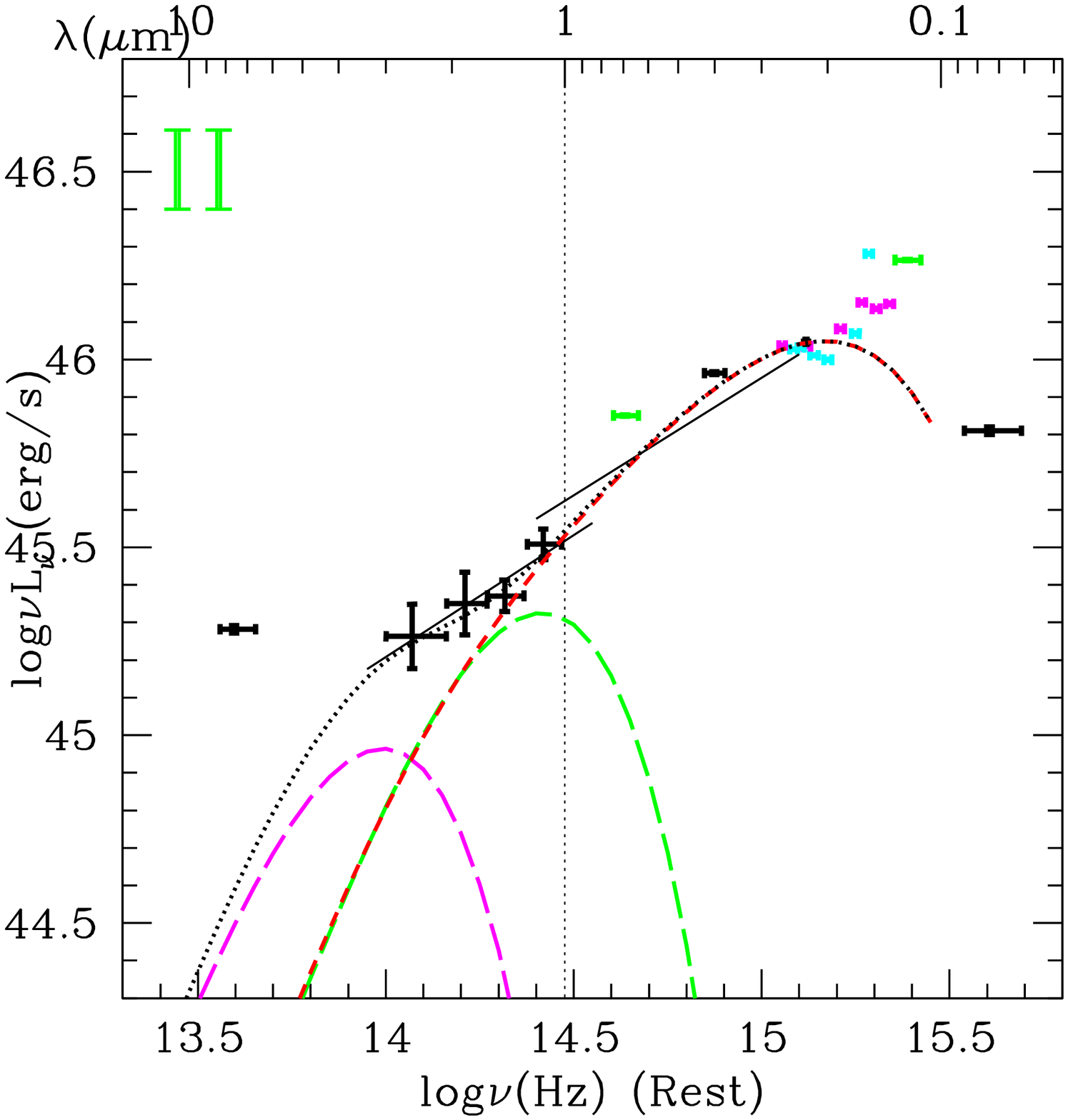}
\includegraphics[angle=0,width=0.325\textwidth]{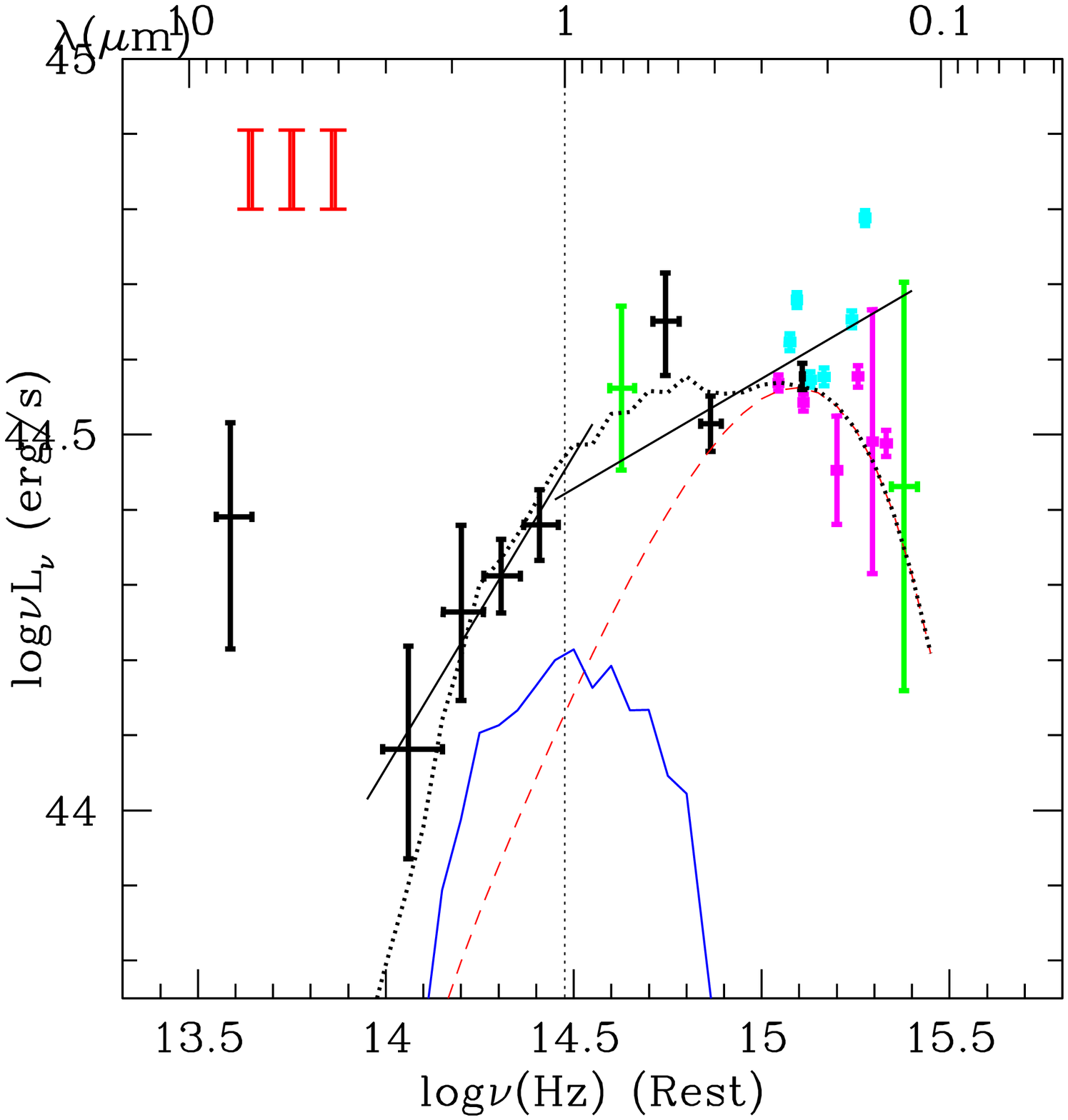}
\caption{Mean SED and examples of SEDs for each HDP classes: Upper
row: mean SED of the three classes of HDP AGNs (black solid line)
compared with E94 RQ mean SED (dotted red line) and the mean SED of
the normal XMM-COSMOS type 1 AGNs with similar BBB (solid red line).
Lower left: class I SED example (XID=2105, redshift $z=1.509$);
lower middle: class II SED example (XID=96, $z=2.117$); lower right:
class III SED example (XID=167, $z=2.048$). The data points are from
Capak et al. (2007), corrected for line emission as in Elvis et al.
(2010). The SEDs are fitted with an accretion disk (red dashed
line), a dust component (magenta dashed line, single temperature
blackbody), and a galaxy component (blue solid line, 5 Gyr
elliptical galaxy). The black dotted line shows the SED of the sum
of the components. The black straight line shows the power-law fits
that give $\alpha_{OPT}$ and $\alpha_{NIR}$. The host galaxy
contribution is negligible, for XID=2105 and XID=96 at high
luminosity (4$L_*$ and 13$L_*$ at 1$\mu$m). The green dashed line
shows the blackbody fitting to outer edge of the accretion disk.
\label{egseds}}
\end{figure*}

Class I. Twenty-four sources lie below the equal slope line (blue
symbols in Figure \ref{slpsel}). These sources have a normal big
blue bump (BBB), but a factor of $\sim$2--4 lower than the E94 mean
SED at 3$\mu$m. They all have a rather flat infrared SED shape
($\alpha_{NIR} \sim 0$) extending to at least $10\mu$m (rest frame;
Figure~\ref{egseds}, left). Severn have high luminosities ($>5 L_*$,
ranging from 5.3$L_*$ to 24$L_*$, where $L_*$ is the break point of
the galaxy luminosity function; Cirasuolo et al. 2007) at 1$\mu$m
and hence should have weak host galaxy contribution, while the rest
have luminosities $0.76 L_*$--$4 L_*$.

Class II. Six sources are consistent with lying on the equal slope
line (green symbols in Figure \ref{slpsel}). The infrared emission
could be the exposed continuation of the BBB to longer wavelengths
($\sim$2--3$\mu$m). Two out of six, have high luminosities ($>5
L_*$, $5.4 L_*$--$13L_*$) at 1$\mu$m, while the rest have
luminosities $0.32 L_*$--$2.7L_*$.

Class III. Eleven sources lie above the equal slope line (red
symbols in Figure \ref{slpsel}). These sources have flatter BBB than
normal, possibly due to reddening. Some of them show quite strong
$\sim 10 \mu$m continuum emission (Figure \ref{egseds}, right). The
class III sources have luminosities of $0.53 L_*$--$5.3L_*$.

We calculated the mean SEDs of the HDP AGNs in each class and
compared them with the mean SEDs of the other type 1 AGNs in the
XMM-COSMOS sample having the same range of optical slopes for that
class, i.e., normal type 1 AGNs with same BBB shape (Figure
\ref{egseds}, upper). The HDP AGNs show relatively weaker 1--3$\mu$m
emission, by a factor of 1.6 for class I, 3 for class II, and 2.5
for class III.

\section{Source Properties}
\subsection{Correlations}
We compared a number of observed properties of the HDP AGNs with the
whole XMM-COSMOS sample: redshift ($z$), bolometric luminosity
($L_{bol}$, 24 $\mu$m--912 \AA), BH mass ($M_{BH}$; Trump et al.
2009b; Merloni et al. 2010), Eddington ratio ($\lambda_E$; Lusso et
al. 2010), the optical to X-ray spectral index
($\alpha_{ox}$)\footnote{$\alpha_{ox}=0.384\times$
log$\left[\frac{F_\nu(2~keV)}{F_\nu(2500\AA)}\right]$,
where $\alpha_{ox}<0$, and larger $\alpha_{ox}$ means 
X-ray louder (Tananbaum et al. 1979).}, the X-ray luminosity at rest
frame 2keV ($L_{2keV}$), the X-ray hardness ratio (HR)\footnote{HR=
$(H-S)/(H+S)$ where $H$ are the XMM counts in the 2--10 keV band and
$S$ those in the 0.5--2 keV band (Brusa et al. 2007, 2010).}, and
radio loudness ($q_{24}$)\footnote{$q_{24}=$log$(f_{24\mu
m}/f_{1.4GHz})<0$ as radio loud (Appleton et al. 2004).}. All the
values of these parameters were reported in Lusso et al. (2010),
Brusa et al. (2010), and Hao et al. (2010). In the HDP sample, only
one source \footnote{XID=167, class III.} is radio-loud (Elvis et
al. 2010).

\begin{figure}
\epsfig{file=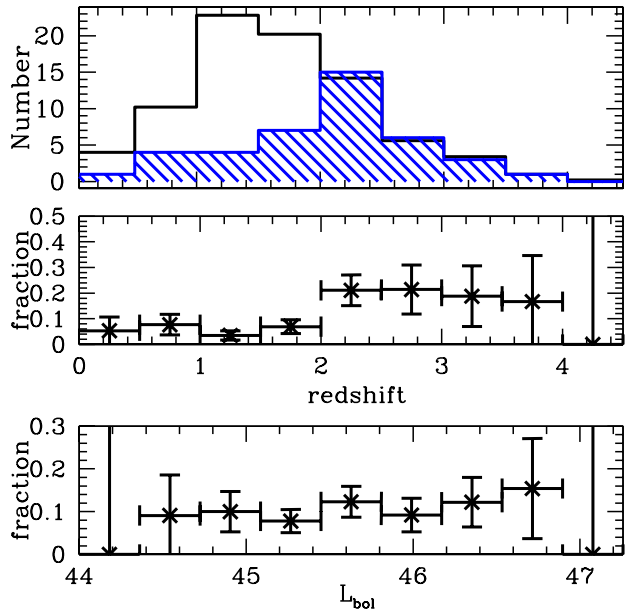, angle=0,width=\linewidth} \caption{Distribution
in the redshift space (top) and fraction of the HDP in the whole
sample as a function of redshift (middle) and $L_{bol}$ (bottom).
The black (number divided by 5) and blue histograms show the
distribution of all the AGNs and the HDP, respectively.
\label{fracbplot}}
\end{figure}

We used the Kolmogorov--Smirnov (K-S) test to compare the
distributions of each parameter for the HDP AGNs with the whole
sample (HDP AGNs excluded). The K-S probabilities are reported in
Table~1. Only $z$ shows a significant correlation
($p_{K-S}=0.02\%$). Distributions of other parameters ($M_{BH}$,
$L_{2keV}$, log$\lambda_E$, and $L_{bol}$) are indistinguishable for
the two samples. We will study the X-ray properties of the HDP AGNs
in a following paper. We can also see some of these results in the
fraction plot in Figure~\ref{fracbplot}, where we divided two
parameters \footnote{One ($z$) with the lowest $p_{K-S}$ and one
($L_{bol}$) with the highest $p_{K-S}$.} into nine bins and plotted
the fraction of HDP AGNs in each bin.

\begin{deluxetable}{ccc}
\tabletypesize{\scriptsize} \tablecaption{K-S Probability
\label{t:ps}} \tablehead{\colhead{Parameter} & \colhead
{$N$(HDP)/$N$(total$-$HDP)\tablenotemark{1}}
 & \colhead{$p_{K-S}$}} \startdata
$z$ & 41/363 & 0.0002 \\
HR&41/363 & 0.080 \\
log$\lambda_E$&16/206& 0.166 \\
$\alpha_{OX}$ & 41/363 & 0.201\\
$L_{2keV}$&41/363 & 0.202\\
$M_{BH}$ & 16/206 & 0.630\\
$L_{bol}$& 41/363 & 0.729 \\
\enddata
\tablenotetext{1}{The number of sources that have the detection of
the physical parameter in the format of (HDP AGN)/(whole type 1 AGN
sample exclude the HDP AGN).}
\end{deluxetable}

The fraction of the HDP AGNs clearly increases with $z$. This
fraction is $5.9\%\pm0.2\%$ at $z<2$ but increases to
$20.4\%\pm0.1\%$ at $2<z<3.5$. Beyond $z=3.5$, the sample is too
small (one source in two bins) to give a useful constraint. A linear
regression fit of the fractions gives
$f_{HDP}=(0.031\pm0.002)(1+z)^{(1.2\pm0.1)}$. The photometry used in
the SED comes from a point source correction to the 3\asec aperture
photometry. Low-redshift SEDs therefore include less host compared
to the high redshift ones. This effect will only move the source
along the mixing curve, so it cannot explain the observed HDP
fraction increase versus redshift.

\subsection{SED Fitting and Physical Parameters}
Studies of NIR spectrum of quasars showed that a blackbody spectrum
is needed to fit the observed spectrum (Glikman et al. 2006). The
BBB is mainly caused by the emission of the accretion disk (e.g.,
Elvis et al. 1994) and could be reddened by the dust in the host
(e.g., Hopkins et al. 2004). At $\sim 1 \mu$m, host galaxy has the
maximum contribution to the observed type 1 AGN SED (e.g., Polletta
et al. 2007). Considering all these effects, we fit all the HDP AGNs
observed NIR--optical SEDs with three components: (1) accretion disk
emission: we use the standard $\alpha$-disk model in Schwarzschild
geometry with electron scattering and Comptonization of soft photons
in the disk atmosphere (Siemiginowska et al. 1995), with measured BH
mass and accretion rate where available (Hao et al. 2010) and
$E(B-V)=$0.1--0.2 in eight cases to get good chi-square in the UV.
(2)hot dust emission: we use a single temperature blackbody
spectrum. (3)host galaxy emission: we use a 5 Gyr old elliptical
galaxy (Polletta et al. 2007). Examples of the fit results are shown
in the lower panel of Figure \ref{egseds}.

\subsubsection{Covering Factor}
The NIR SEDs are fitted with a blackbody to get the maximum dust
temperature ($T_d$). The sublimation temperature for graphite and
silicate grains are $\sim$ 1500--1900 K and $\sim$ 1000--1400 K,
respectively (e.g., Salpeter 1977; Laor \& Draine 1993). $T_d$
ranges from 800 K to 1900 K. Most (19 out of 34) HDP AGNs with dust
component have $T_d>$1400~K, suggesting graphite grains.

Dust can only exist at a radius beyond the dust evaporation radius,
given by $r_d=1.3L_{uv, 46}^{1/2}T_{1500}^{-2.8}$ pc (Barvainis
1987), where $L_{uv,46}$ is the total ultraviolet (1$\mu$m--912\AA)
luminosity in units of $10^{46}$ erg s$^{-1}$ and $T_{1500}$ is the
maximum dust temperature in units of 1500~K. We find $r_d$ ranges
from 0.2~pc to 3.9~pc for the HDP sources. The emission area ($A_d$)
is given from the blackbody fitting, and $A_d$ ranges from 0.1
pc$^2$ to 10.68 pc$^2$. The covering factor of the dust component is
then $f_c=A_d/(4\pi r_d^2$). For example, the results for source
XID=2105 (Figure \ref{egseds}, left) are $T_d$=1500~K and
$A_d$=1.76~pc$^2$. The evaporation radius is 0.83~pc, corresponding
to $f_c$= 20\%. $f_c$ ranges from 1.9\% to 29\%, which are a factor
of 2--40 smaller than the 75\% expected from the unified models to
give the observed type 2 to type 1 ratio (e.g., Krolik \& Begelman
1988). Even if considering the dependence of type 2 to type 1 ratio
on X-ray luminosity (e.g., Gilli et al. 2007), the $f_c$ of HDP AGN
is still smaller than the 50\% expected for the X-ray bright AGNs.

The above innermost radius estimation is comparable to those
reported in Kishimoto et al. (2007, 2009). 
Kishimoto et al. (2007) also showed that the innermost radii
measured by near-IR reverberation are systematically smaller by a
factor of $\sim$ 3 than the value predicted with the Barvainis
(1987) formula. For the HDP AGNs, if the innermost radius of the hot
dust is really a factor of 3 smaller, the covering factor would be a
factor of 9 larger than the values we calculated, which could reach
the 75\% required by the unified models (e.g., Krolik \& Begelman
1988) for most (29 out of 34) of the HDP sources.

\subsubsection{Disk Outer Radius}
For 11 HDP AGNs with little host galaxy contamination (defined as
$<$50\% at $1\mu$m), and available $M_{BH}$ and $\lambda_E$ (Lusso
et al. 2010; Trump et al. 2009b; Merloni et al. 2010), the lack of
hot dust emission allows us to see what appears to be the accretion
disk emission extending further out well beyond $1\mu$m.

We fitted the outer edge of the accretion disk component (red dashed
line in Figure \ref{egseds}) with a single temperature ($T_c$)
blackbody spectrum (green dashed line). The outer radius of the
accretion disk (Frank et al. 2002):
$$R_{out}=1.1\times 10^4 T_c^{-\frac43}\alpha^{-\frac{4}{15}}
\eta^{-\frac25}M_8^{\frac{11}{15}}\lambda_E^{\frac25}f^{\frac85}~~~pc,$$
where $M_8=M/(10^8 M_{\bigodot} )$ and $f=\left [1-\left
(\frac{6GM}{R~c^2}\right )^{\frac{1}{2}}\right ]^{\frac{1}{4}}$. We
assume $\alpha=0.1$, $\eta=0.1$.

An example is the source XID=96 shown in Figure \ref{egseds} (lower
middle; $M_8=3.55$, $\lambda_E=0.71$), the blackbody temperature at
outer edge is 3200~K. This gives $R_{out}=0.47$ pc, $\sim 1.3\times
10^4$ Schwarzschild radius ($r_s$), 13.6 times the gravitational
stability radius of the accretion disk (Goodman 2003, i.e., the
radius beyond which the disk is unstable to self-gravity and should
break up). For the 11 HDP AGNs, the $T_c$ range from 2200~K to
4500~K. The $R_{out}$ range from 0.09~pc to 0.99~pc, corresponding
to $(0.29-2)\times10^4 r_s$. These $R_{out}$ are $\sim$10--23 times
larger than the gravitational stability radius.

\section{Discussion and Conclusions}
In a sample of 404 XMM-selected type 1 AGNs (excluding four with
incomplete NIR photometry), we found that $10.1\%\pm 1.7\%$ have
weak NIR emission, indicating a relative paucity of hot dust
emission. We call these HDP AGNs.

HDP AGNs have not been reported from SDSS or earlier samples. We
have made an initial check of where the Richards et al. (2006) and
E94 samples lie on the $\alpha_{OPT}-\alpha_{NIR}$ plane. We find
similar fraction of HDP AGNs 
and will report on this study in detail in a later paper. The
$\alpha_{OPT}-\alpha_{NIR}$ plane is a useful tool for locating
outliers.

The fraction of HDP AGNs clearly increases to redshift $z=3.5$
($f_{HDP}=(0.031\pm0.002)(1+z)^{(1.2\pm0.1)}$). No trends of the
fraction of HDP AGNs with other parameters, notably $L_{bol}$, are
observed.

We divided these HDP AGNs into three classes according to their
$\alpha_{OPT}$ and $\alpha_{NIR}$. We fitted the HDP SEDs with three
components: the accretion disk, a blackbody to represent the hot
dust, and the host galaxy. We found the dust covering factors are
1.9\%--29\%, well below the typical 75\% required by unified models
(e.g., Krolik \& Begelman 1988). For the 11 HDP AGNs with little
host contribution ($<$ 50\% at 1 $\mu$m) and available $M_{BH}$ and
$\lambda_E$, we estimated the lower limit of the accretion disk
outer radius to be (0.29--2)$\times10^4 r_s$ (0.09--0.99~pc),
corresponding to 10--23 times the gravitational stability radius of
an $\alpha$ disk. How these disks stabilized is an open question, or
the long wavelength turndown is not $R_{out}$ as we assumed. These
results agree with the NIR disk spectrum uncovered using polarized
light in Kishimoto et al. (2008). We ignore further discussion of
the sources with strong host galaxy contribution as the system is
more complicated to include the also unknown galaxy part.

There are several possibilities to explain the lack of NIR emission
in HDP sources. First, most  XMM--COSMOS HDPs are at $1.5<z<3$, when
the universe is 2.1--4.2 Gyr old, and they have 1--3 Gyr from
reionization to form a torus. The Jiang et al. (2010) proposed
explanation thus seems unlikely for the XMM--COSMOS HDP AGNs.
Second, the luminosity and Eddington ratio distributions of the HDP
AGNs and the normal XMM--COSMOS type 1 AGNs are indistinguishable.
This rules out the possibility that HDP AGNs are low-luminosity or
low-accretion rate sources, unable to support a dusty torus (Elitzur
\& Ho 2009). Third, evolutionary scenarios (e.g., Hopkins et al.
2008) predict short-lived Eddington limited outbursts after a
merger, which could destroy the innermost dust. The HDP AGNs might
then be quasars that had insufficient time to reform dust in the
inner region. Fourth, alternatively, when two galaxies merge 
, the chaotic dynamics might destroy the ``torus''. The `torus' then
reforms in a timescale that might allow 10\% of quasars to be HDP.
The dust formation timescale depends strongly on pressure and
temperature (Whittet 2003; Kr\"{u}gel 2003) making estimates
unreliable as both quantities are poorly known. Last, Guedes et al.
(2010) suggested that a BH ejected by gravitational wave recoil and
carrying along its accretion disk and broad emission line region,
but not the hot dust ``torus'', would appear as an HDP AGN.

The unusual SEDs of HDP AGNs can affect estimates of galaxy bulge
and BH masses derived from SED fitting assuming the BH SED to be E94
mean SED. Merloni et al (2010) recently used this method to show
that $z\sim1.5$ quasars deviate from the local $M-\sigma$
relationship (H\"{a}ring \& Rix 2004). Three of the Merloni et al.
(2010) quasars belong to our HDP sample. For these sources, the SED
fitting using E94 template gives a small AGN contribution. Allowing
an HDP SED to fit would increase the BH mass and decrease the
stellar mass, making them deviate even more from the local
$M-\sigma$ relationship. New AGN templates with variable dust bump
strengths are needed to derive accurate galaxy and BH masses in
these objects.

The properties of the HDP AGNs still need to be investigated to
understand their formation scenario and put them in the context of
galaxy and SMBH evolution. In particular, we will check the existing
Hubble images and optical and X-ray spectra of HDP AGNs in future
papers.

\section{Acknowledgments}
H.H. thanks Sumin Tang for useful discussions. This work was
supported in part by NASA Chandra grant number GO7-8136A (H.H.,
F.C., M.E.). In Italy this work is supported by ASI/INAF grants
I/023/05.

\end{document}